\documentclass[aps,pra,twocolumn,groupedaddress,showpacs,showkeys]{revtex4}
\usepackage{graphicx}
\usepackage{amssymb}
\usepackage{bm}

\begin{document}

\title{Bose-Einstein Condensates in Optical Lattices: Band-Gap Structure and Solitons}

\author{Pearl J.Y. Louis$^1$, Elena A. Ostrovskaya$^{1,2}$, Craig M. Savage$^2$, and Yuri S. Kivshar$^1$}

\affiliation{$^1$Nonlinear Physics Group, Research School of
Physical Sciences and Engineering, Australian National University,
Canberra ACT 0200, Australia \\
$^2$Department of Physics and Theoretical Physics, Australian
National University, ACT 0200, Australia}

\begin{abstract}
We analyze the existence and stability of spatially extended
(Bloch-type) and localized states of a Bose-Einstein
condensate loaded into an optical lattice. In the framework of the
 Gross-Pitaevskii equation with a periodic
potential, we study the band-gap structure of the matter-wave
spectrum in both the linear and nonlinear regimes. We demonstrate the
existence of families of spatially localized matter-wave gap solitons, and
analyze their stability in different band gaps, for both repulsive
and attractive atomic interactions.
\end{abstract}

\maketitle

\section{Introduction}

Trapping of neutral atoms in optical lattices \cite{Sal_et87}
provides an effective and powerful means for controlling the
properties of Bose-Einstein condensates (BECs), as confirmed by a
number of the recent experiments (see, e.g., Refs.
\cite{And_Kas98,Orz_et01,Bur_et01,Cat_et01,Arimondo1_02,Arimondo2_02,Phillips_02}).
Coupling between the BEC droplets localized in weakly interacting
minima of a standing light wave creates a reconfigurable matter-wave structure
that can be easily manipulated by varying the strength of the
periodic potential and its well spacing.

Many properties of such arrays of the BEC droplets can be
understood, within the framework of the mean-field approach,  by employing the  tight-binding approximation borrowed from solid state physics \cite{Jav99,Tro_Sne01,Abd_et01}. Being based on the assumption of strong localization of the
condensate wave functions in the individual potential wells of the lattice, the tight-binding approximation and the resulting discrete lattice models provide a limited description of the BEC dynamics in an optical lattice. An alternative reduction of the continuous system to the discrete vector lattice involves employing a complete set of on-site Wannier states \cite{wannier_02}, and is more general than a tight-binding approximation. However, it is to be clarified whether such a reduction can provide predictions on the dynamics and stability of a BEC in a periodic potential more readily than the full mean-field model. A more adequate analysis of both periodic and localized modes of the continuous Gross-Pitaevskii (GP) equation with a periodic lattice potential is still missing. Some of the recent studies of the GP model \cite{Zob_et99,Bro_et01,Kon_Sal02,Alfimov_02} point towards
interesting features of the BEC dynamics that should be observed
beyond the applicability of discrete models.

The main purpose of this paper is to analyze the structure and
stability of the stationary states of a
condensate in an optical lattice, the so-called nonlinear modes \cite{Kiv_et01}, 
in the framework of a continuous
mean-field model. This approach allows us to consider more general
types of stationary states and BEC dynamics without many
of the restrictions of the discrete models. We solve the continuous
one-dimensional GP equation numerically and describe families of
stationary states, which possess different spatial structure and
stability properties.

First, we find spatially extended stationary {\em
nonlinear periodic modes} of the condensate in a lattice. These
periodic modes possess different symmetry, and provide a
generalization of both the Bloch modes in the
theory of linear periodic systems and of the extended states of a condensate in a finite optical lattice with a confining potential \cite{Weiping_98}. We demonstrate that the matter-wave spectrum for such periodic BEC states has a band-gap structure, which survives in the presence of the mean-field nonlinearity in the GP equation. 

We also find novel types of condensate states which cannot exist in the linear limit. These are {\em  spatially localized
modes} of the condensate which exist in the band gaps of the
matter-wave spectrum and, therefore, they should be called {\em
gap solitons}. The existence of the matter-wave gap solitons was first suggested in the framework of the coupled-mode theory \cite{Zob_et99}, in analogy to optical gap solitons in Bragg gratings. Remarkably, gap solitons - spatially localized atomic wavepackets - may exist even in a condensate with {\em repulsive} interatomic interactions  \cite{Zob_et99}.  Their experimental demonstration, however, remains a challenge. Here, we present a comprehensive theory of the structure and linear stability analysis of {\em different families
of gap solitons} with respect to variations in the lattice
parameters and effective nonlinearity defined by the number of
atoms in the condensate. We discuss the key scenarios of
instability of spatially localized modes in optical lattices and
the physical mechanism for creating different multiple
states of the BEC droplets.

We would like to emphasize that many of the effects described here
in the framework of the GP equation with a periodic potential,
such as the mode stability and the existence of the localized modes in different band gaps, cannot be obtained in the framework of the tight-binding approximation since the coherent
and spectral properties of these two models are different. This latter issue is
similar to the BEC dynamics in a double-well potential where a
rigorous coupled-mode wave theory should be based on the analysis
of nonlinear modes of the entire structure, rather than 
condensate wave functions of isolated wells \cite{Ost_et00}.

The paper is organized as follows. In Sec. II we present our model
and reduce it to the one-dimensional GP equation, under the
assumption of a cigar-shaped condensate. Section III includes the
analysis of both linear and nonlinear matter-wave spectra in an
optical lattice. We revisit the structure of the band-gap
spectrum for the Bloch-like modes of the noninteracting condensate, and extend
this concept to the nonlinear case, demonstrating the existence of
the nonlinear Bloch-like modes and the nonlinearity-dependent
band-gap structure of the matter-wave spectrum in optical
lattices. The study of the BEC gap solitons in different band gaps is summarized in Sec.
IV, where we also present our results on the stability of gap
solitons, for both repulsive and attractive atomic interactions. 
Section V concludes with a summary and further
perspectives.

\section{Model}

We consider the macroscopic dynamics of a BEC loaded onto an
elongated optical lattice, similar to those created in a number of
experiments (see, e.g., Refs. \cite{Sal_et87,And_Kas98,Orz_et01}).
The condensate dynamics is described by the three-dimensional GP
equation,
\[
i\hbar\frac{\partial\Psi}{\partial T} =
\left[-\frac{\hbar^{2}}{2m}\nabla^{2} + V(X,Y,Z) +
\Gamma |\Psi|^{2}\right]\Psi,
\]
where $\Psi(X,Y,Z;T)$ is the wave function of the cigar-shaped
condensate, $(Y,Z)$ are the directions of strong transverse confinement,  and $X$ is the direction of the lattice. The combined potential of the optical lattice and magnetic trap $V(X,Y,Z)$ can be written as, $V(X,R^2)=E_0 \sin^{2}(\pi X/d) + (1/2)m({\omega_x}^2 X^2+\omega^2_\perp R^2)$,  where $R^2=Y^2+Z^2$, $E_0$ is the well depth of the optical lattice, $d$ is the characteristic lattice constant, and $\omega_i$ are trapping frequencies in the corresponding directions.  The
parameter $\Gamma=4\pi\hbar^{2}a_{s}/m$ characterizes the s-wave
scattering of atoms in the condensate which introduces an
effective nonlinearity in the mean-field equation; it is positive for repulsive
interactions and negative for attractive interactions.

Measuring time in units of $\omega^{-1}_\perp$, the spatial variables in units of the transverse
harmonic oscillator length,  $a_0=(\hbar/m\omega_\perp)^{1/2}$, the
wave function amplitude in units of $a^{-3/2}_0$,
and the potential in units of $\hbar\omega_\perp$, we obtain the
following dimensionless GP equation:
\begin{equation}
\label{dimensionlessGPE} i\frac{\partial\Psi}{\partial t} = \left[
-\frac{1}{2}\nabla^{2} + V(x,r^2) + g_{\rm 3D}|\Psi|^{2} \right]
\Psi,
\end{equation}
where $t=T\omega_\perp$, $(x,r)=(X,R)/a_0$, and $g_{\rm 3D}=4\pi(a_s/a_0)$ .  The potential now takes the form $V(x,r^2)=V_0\sin^2(Kx)+(1/2)(\Omega^2 x^2+r^2)$,where $V_0=E_0/(\hbar \omega_\perp)$, and $K=\pi a_0/d$. The ratio of the confinement strengths for the magnetic trap $\Omega=\omega_x/\omega_\perp$ varies from $~10^{-1}$  \cite{Esslinger_01,Bur_et01,Cat_et01,Orz_et01} to $1/\sqrt{2}$ \cite{Arimondo1_02,Arimondo2_02,Phillips_02}.

To simplify our analysis, we consider a cigar-shaped condensate in a strongly elongated trap ($\Omega \sim 10^{-1}$) and a quasi-one-dimensional optical lattice in the direction of a weak confinement. Under these assumptions, the model can be reduced to a one-dimensional GP equation by
assuming separable solutions of the form,
$\Psi(r,x;t)=\Phi(r)\psi(x;t)$. Due to the tight confinement of the condensate in the transverse
direction, we assume that the spatial structure of the function $\Phi(r)$ is well described by a solution of the two-dimensional radially symmetric quantum harmonic
oscillator problem, $\nabla^{2}_{\perp}\Phi + \Phi - r^{2}\Phi = 0$.
Taking the ground state solution, $\Phi(r)=C \exp(-r^{2}/2)$,
where $C^{2}=1/\pi$ is found from the normalization condition,
$\int_{-\infty}^{\infty} |\Phi|^{2} dy dz = 1$, and integrating
Eq. (\ref{dimensionlessGPE}) over the transverse coordinate, we
derive an effective one-dimensional GP equation in the
dimensionless variables,
\begin{equation}\label{dimless}
\label{eq1DGPE} i\frac{\partial\psi}{\partial t} =
-\frac{1}{2}\frac{\partial^{2}\psi}{\partial x^2} + V(x)\psi +
\sigma g_{\rm 1D} |\psi|^{2}\psi,
\end{equation}
where $g_{\rm 1D}=2|a_s|/a_0$, and the
coefficient $\sigma=\text{sgn}(a_{s})=\pm 1$ characterizes the type of the
s-wave interactions. Due to the weak magnetic confinement in $x$ direction, it is possible to omit the term $~\Omega^2$ from the effective potential, thus assuming that the 1D lattice potential , $V(x)=V_0\sin^2(Kx)$, is uniform and is not affected by the presence of the additional confining potential. Furthermore, it is convenient to reduce the parameter space of the model i.e. effectively set $g_{\rm 1D}=1$) by rescaling the wave function as $\psi \to \psi \sqrt{g_{\rm 1D}}$. In physical terms, it means that the peak density of the condensate is used to characterize the effective mean-field nonlinearity.  

The model  (\ref{dimless}) can describe the BEC in a one-dimensional optical lattice over a wide range of experimental parameters, such as lattice depth, well spacing, and the effective nonlinearity. Indeed, Eq. (\ref{dimless}) is invariant with respect to the following scaling transformation: $\{ K,V_0,x,t,\psi\}\to\{K/\eta, V_0/\eta^2,x\eta,t\eta^2,\psi/\eta\}$, where $\eta$ is a dimensionless constant. However, a significant point of difference
between the lattice in our model and those in
experiments~\cite{Sal_et87,And_Kas98,Orz_et01} is the absence of
an additional confining potential (e.g., gravity or magnetic trapping).

We determine the experimentally relevant range of dimensionless parameters for our model by taking the case of  $^{87}\text{Rb}$ atoms, which sets $m=1.44\times 10^{-25}$ ${\rm kg}$ and
$a_{s}=5.3$ ${\rm nm}$. The characteristic lattice constant, $d$, is determined by the angle between the intersecting laser beams forming the lattice and, in the current experiments,  can be varied in the range $0.4 - 1.6$ $\mu{\rm m}$ \cite{Arimondo1_02,Arimondo2_02}. The lattice depth, $E_0$, scales linearly with the light intensity, and varies between zero and  $E^{\rm max}_0 \sim 20E_{\rm rec}$, where $E_{\rm rec}=\hbar^2\pi^2/(2md^2)$ is the lattice recoil energy \cite{Arimondo1_02,Arimondo2_02}. Taking the transverse frequency as $\omega_\perp \sim 2\pi \times 10^2$ ${\rm Hz}$ \cite{Orz_et01,Bur_et01},  we obtain the following regions of the model parameters: $2 \leqslant K \leqslant 8$ and $4.5 \times 10^1 \leqslant V^{\rm max}_0 \leqslant 7.0 \times 10^2$, respectively. In what follows, for the sake of computational convenience, we will make use of the scaling properties of the model and use two different sets of the parameters: $K=0.1, V_0 \sim 0.1$ ($\eta=5$), and $K=0.4, V_0 \sim 1.0$ ($\eta=20$). 

We note that the number of atoms in the condensate wavefunction, ${\cal N}$,  is now given by the expression
${\cal N}=N/g_{\rm 1D}$,  where
\begin{equation}\label{Neq}
N=\int_{-\infty}^{\infty} |\psi(x,t)|^{2} dx
\end{equation}
is an integral of motion for the model (\ref{eq1DGPE}).

\section{Band-Gap Spectrum of Matter Waves}

\subsection{Linear Bloch modes}

Stationary states of the condensate in an infinite periodic potential of an optical lattice
(`nonlinear modes') are described by solutions of Eq.
(\ref{eq1DGPE}) of the form: $\psi (x,t) = \phi(x) \exp (-i\mu t)$,
where $\mu$ is the corresponding chemical potential. The
steady-state wave function $\phi (x)$ obeys the {\em
time-independent} GP equation:
\begin{equation}
\label{eqtindepGPE} \frac{1}{2}\frac{d^{2} \phi }{dx^{2}}
-\left[ V_0\sin^2(Kx) \phi-\mu\phi \right] - \sigma |\phi|^{2}\phi=0,
\end{equation}
where the lattice depth $V_0$ and period $K$ are the
experimentally controlled parameters, and they both can be varied
in our model.

In the case of noninteracting condensate, which formally corresponds to $\sigma=0$, Eq.
(\ref{eqtindepGPE}) is linear in $\phi$ and, as has been shown in \cite{bands_98}, the Bloch-function formalism applies. In this case (\ref{eqtindepGPE}) can be written as the
Mathieu's equation:
\begin{equation}\label{math}
\label{mathieu} \frac{d^{2}\phi }{dy^{2}}-[ 2q\cos(2y)-p]\phi=0,
\end{equation}
where $y=Kx$, $q=-V_0/(2K^2)$, and $p=2(q+\mu/K^2)$. By employing
the Floquet-Bloch theorem, the condensate wave function can be
presented as a superposition of {\em Bloch waves},
\begin{equation}
\phi(y)=b_1P_1(y) e^{i\lambda y} + b_2P_2(y) e^{-i\lambda y},
\end{equation}
where $P_{1,2}(y)$ are periodic functions with the least period
$\pi$, $b_{1,2}$ are constants, and $\lambda$ is the
characteristic of the Floquet exponent. The structure of all
solutions to the Mathieu's equation can be determined by Floquet
theory (this text-book analysis can be found, e.g., in Ref.
\cite{mathieu}). According to this theory, the spectrum of Eq. (\ref{math}) consists of domains   (`bands') of the $(p,q)$ space in which there exist only amplitude-bounded, oscillatory solutions. The bands are separated
by the regions (`gaps') where unbounded solutions exist. The
band edges correspond to exactly periodic solutions.

\begin{figure}[tbp]
\includegraphics[width=7.5cm]{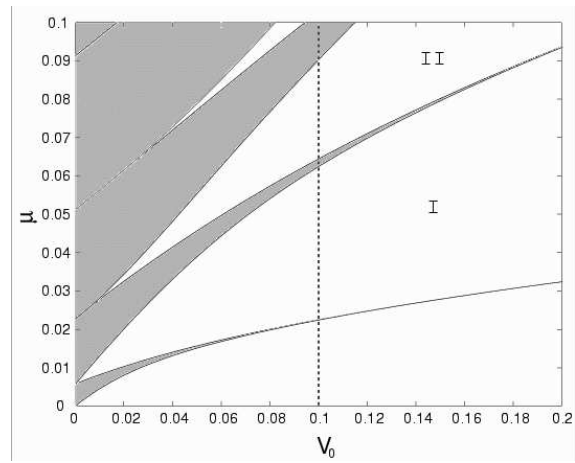}
\caption{\label{fig_stability} Band-gap diagram for Bloch waves in the linear regime 
(noninteracting BEC in an optical lattice).
Shaded areas: bands of bounded oscillatory solutions for the
matter waves; numbered areas (I and
II)- the lowest two band gaps in the spectrum where some
solutions are unbounded and no periodic solutions exist. Solid
lines - band edges corresponding to periodic solutions.
Dashed line shows the particular value $V_0=0.1$ used below.}
\end{figure}

Figure \ref{fig_stability} presents the band-gap diagram for the
extended solutions of Eq. (\ref{mathieu}) which describe noninteracting
condensed atoms in an optical lattice. The results are presented
for the parameter domain $(\mu, V_0)$ relevant to our problem. The
shaded areas correspond to the regions where the Floquet exponent
has a real part (which is an odd or even integer) and, therefore,
spatially oscillating (but not in general periodic) solutions to the
linear model equation (\ref{eqtindepGPE}) exist.  If $\lambda$
has a nonvanishing imaginary part, the solutions do
not oscillate (decay or grow exponentially with $x$); such solutions belong to the gaps separating the bands of oscillatory solutions. The first two gaps where
unbounded solutions exist, numbered $\rm{I}$ and $\rm{II}$ in Fig.
\ref{fig_stability} are bordered by the band edges representing the
linear {\em Bloch waves} with periodicity of $2\pi/K$ and $\pi/K$,
respectively.

It is in these gaps of the linear spectrum that the nonlinear localization of matter waves in the form
of  {\em gap solitons} occurs, and it is this localization that could
enable the creation of solitons in the BEC with repulsive
interatomic interaction. The chemical potentials corresponding to
such localized waves lie in the band gaps of the spectrum of the noninteracting condensate  (marked areas in
Fig. \ref{fig_stability}). 

The dashed line in Fig. \ref{fig_stability} marks the value of the lattice depth for which the the study of the gap modes presented below is conducted. We note that this value is greater than the threshold value $V^* = \mu$ that roughly marks the transition between the quasi-unbounded ($V_0<V^*$) and strongly bounded ($V_0 > V^*$) condensate wavefunctions (see also discussion in Ref. \cite{Drese}). The tight-binding or "fragmented" condensate regime corresponds to $V_0 \gg V^*$, where the bands "collapse" to discrete levels of the Wannier states \cite{bands_98}.

The knowledge of a complete band-gap spectrum of the matter-waves in an optical lattice provides us with important clues on
the existence and stability regions of different types of BEC 
localized states that may be excited in an optical lattice. The
frequently employed nearest-neighbor tight-binding approximation based on a
discrete model \cite{Tro_Sne01,Tro_Sne-2,Abd_et01} is inferior to
the analysis of the complete continuous model in that it describes
only one (the first) band surrounded by two semi-infinite gaps.
The following sections aim to provide us with understanding of the
details of the formation and stability of "gap modes" - the
nonlinear localized states of the condensate in a lattice. Despite
the recent advances in the treatment of the continuous GP models
\cite{Bro_et01,Wu_02,Alfimov_02}, such an analysis is still missing 
\footnote{At the time of final preparations of this publication a paper containing stability analysis of fundamental matter-wave solitons in the first gap has appeared on Los Alamos preprint archive: K.M. Hilligs{\o}e et al., arXiv:cond-mat/0208136. The results of our analysis are consistent with the findings of this work}.

\begin{figure}[htbp]
\includegraphics[width=7.5cm]{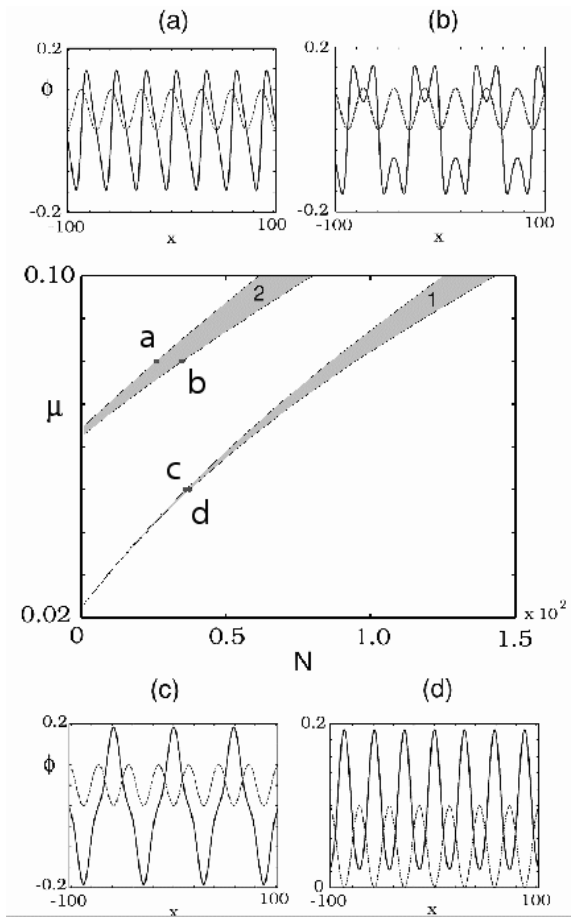}
\caption{\label{pearl_fig0} Middle: Condensate population (shown as the
number of atoms, $N$, defined by Eq. (\ref{Neq}), per lattice period, $\pi/K$) for the
extended stationary states of a repulsive interaction ($\sigma>0$) BEC in an optical lattice at
$V_0=0.1$. The values of $\mu$ corresponding to the linear band
edges at $V_0=0.1$ are recovered in the limit  $N \to 0$ (cf. Fig.
\ref{fig_stability}, dashed line). Top and bottom: dotted lines - structure of the lattice potential, and solid lines - spatial structure of the
nonlinear Bloch waves at the marked points on the band edges.}
\end{figure}

\subsection{Nonlinear Bloch modes}

The effect of interatomic interaction, described by the cubic
nonlinearity in the mean-field GP equation, results in an effective shift of
the chemical potential for the extended periodic states, i.e. Bloch
waves of BEC, on the band edges. This leads to the corresponding
shifts of the band-gap edges, and the gaps can even close up for large
densities of the repulsive condensates. The nonlinearity-induced
shift of the bands can be described by the multi-scale
perturbation theory developed for the Mathieu's equation with
cubic nonlinearity (\ref{eqtindepGPE}) and a shallow potential
($V_0 \ll 1$) (see, e.g. Ref. ~\cite{El-Dib_01}). Here, we study the
extended periodic stationary states of the interacting condensates
in the form of {\em nonlinear Bloch waves}, by solving Eq.
(\ref{eqtindepGPE}) numerically with standard relaxation techniques.

In correspondence to the linear odd and even Bloch waves that
satisfy Eq. (\ref{mathieu}), {\em the nonlinear Bloch waves}
display similar spatial structures, in which the droplets of
the neighboring wells are out of phase or in phase with each other
(see Fig. \ref{pearl_fig0}). The families corresponding to
the two distinct stationary states for different condensate
densities are shown in Fig. \ref{pearl_fig0} for the repulsive
condensate ($\sigma=+1$). The departure of the nonlinear Bloch
states from the linear limit (recovered at $N\to 0$) indicates the 
degree to which the mean-field nonlinearity affects the structure
of the band gaps in the matter-wave spectrum. However large this
departure is, the spatial periodicity and symmetry of the linear
Bloch waves corresponding to different bands (see, e.g., Ref.
\cite{Frenkel_01}) is preserved in the nonlinear regime. The
"in-phase" and "out-of-phase" extended periodic states of the
condensate corresponding to the lower and upper edges of the
lowest band (see Fig. \ref{pearl_fig0}, bottom panel) represent a
continuous-wave analogue to the so-called "unstaggered" and
"staggered" modes of the discrete lattice model, respectively
\cite{Abd_et01}. However, none of the higher-order band modes can
be described by the discrete model, in the framework of the nearest-neighbor tight-binding approximation, and its validity is therefore
restricted to the lowest band only.


\section{Matter-Wave Gap Solitons}

\subsection{Spatially localized modes}

\begin{figure}[h]
\includegraphics[width=8.5cm]{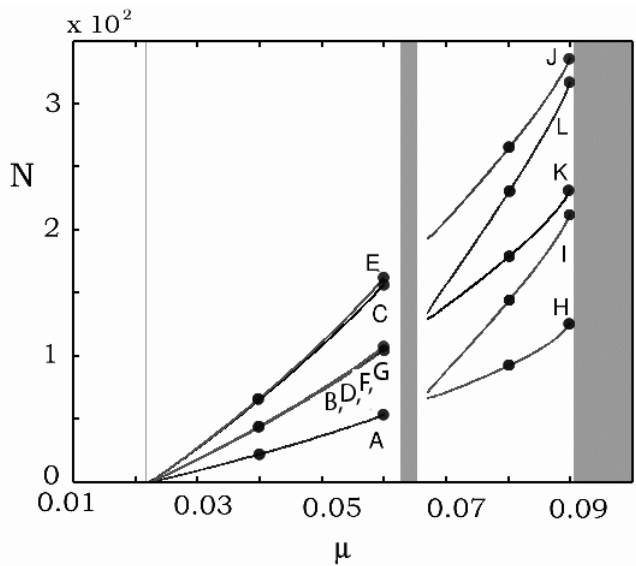}
\caption{\label{fig_families} Families of different gap
solitons. The structure of the modes corresponding to the marked
points is shown in Figs. \ref{fig_profiles1_1} ,  \ref{fig_profiles1_2}, and 
\ref{fig_profiles2}.  Shaded areas show the linear Bloch-wave bands.}
\end{figure}

\begin{figure}[hbp]
\includegraphics[width=7.5cm]{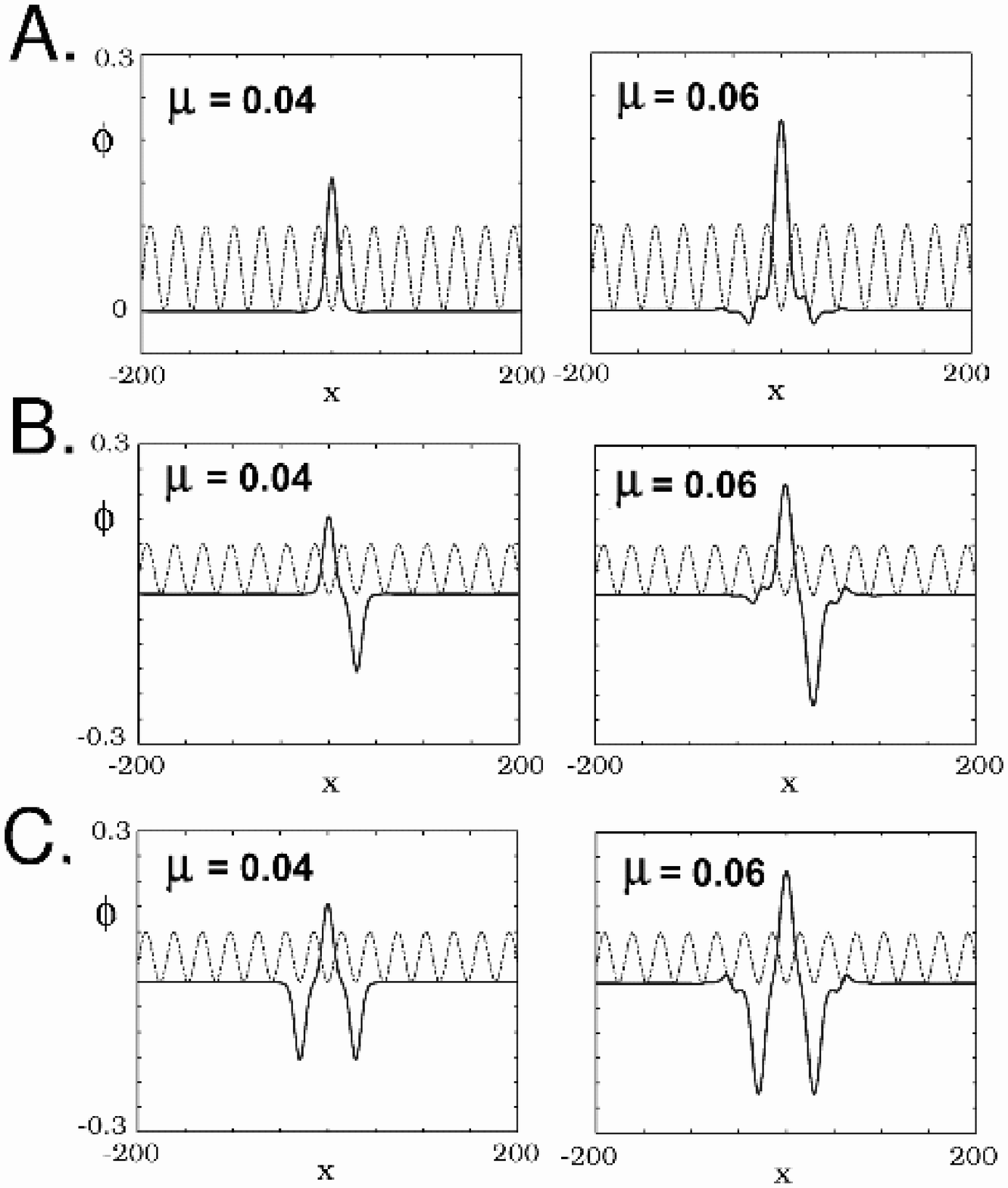}
\caption{\label{fig_profiles1_1} Spatial structure of the localized
states in the first band gap corresponding to the marked families A, B, and C in Fig.
\ref{fig_families}. }
\end{figure}

\begin{figure}[tbp]
\includegraphics[width=7.5cm]{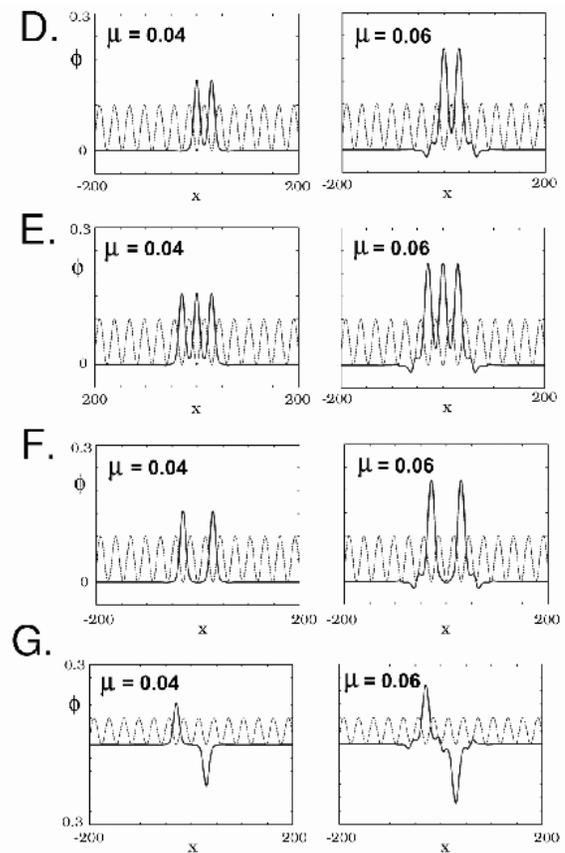}
\caption{\label{fig_profiles1_2} Spatial structure of the localized
states in the first band gap corresponding to the marked families D, E, F, and G in Fig.
\ref{fig_families}. }
\end{figure}

\begin{figure}[tbp]
\includegraphics[width=7.5cm]{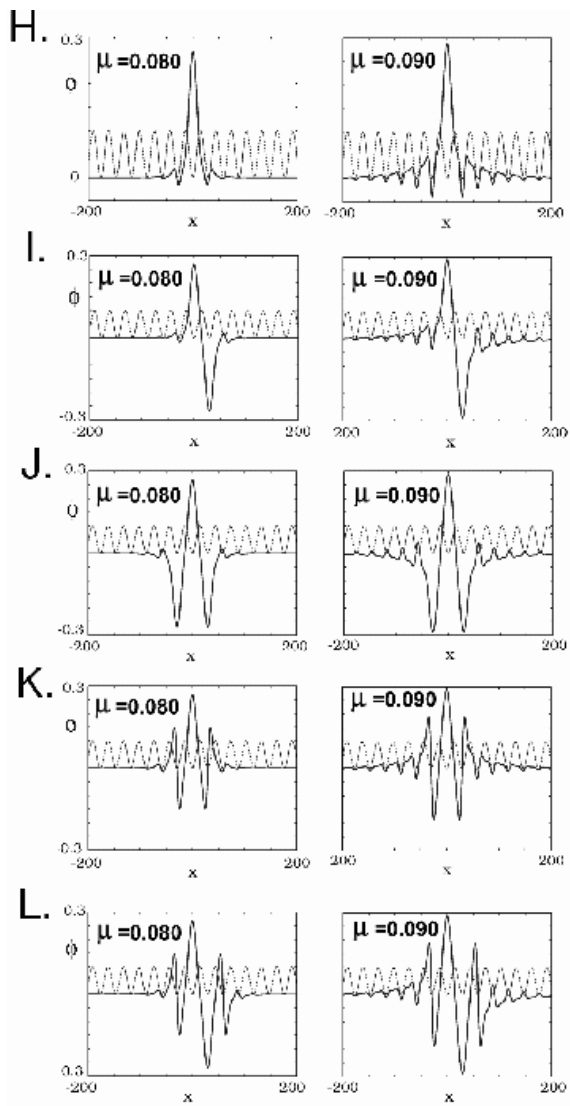}
\caption{\label{fig_profiles2} Spatial structure of the localized
states in the second band gap corresponding to the marked families in Fig.
\ref{fig_families}. }
\end{figure}

\begin{figure}[tbp]
\includegraphics[width=7.5cm]{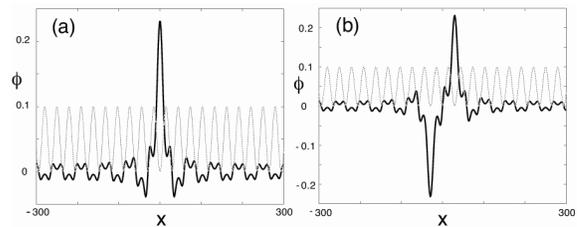}
\caption{\label{fig_profiles3} Spatial structure of the states in
the second transmission band, localized on the extended background
and corresponding to the families (a) A and (b) B  in Fig
\ref{fig_profiles2}. }
\end{figure}

We have numerically found different families of spatially
localized BEC states in an optical lattice, for both repulsive and
attractive interaction. Such states can be regarded as strongly
localized soliton-like modes created in band gaps of the
optical-lattice spectrum, and therefore they should be compared
with gap solitons found for other periodic systems beyond the coupled-mode theory (see, e.g.,
Refs.~ \cite{sukhorukov,sukh_pre} and references therein). Several families
of the lowest-order gap modes and their spatial structure are
presented in Figs. \ref{fig_families} to \ref{fig_profiles3}
for $\sigma = +1$. The gap solitons exist in all band gaps,
excluding the semi-infinite gap of the spectrum below the first
band, which is analogous to the total internal reflection gap in optical Bragg gratings.


One of our main results is that each of the band gaps includes a
branch of the lower-order fundamental gap modes possessing a
single central peak. These fundamental localized states, or {\em
the BEC droplets}, are strongly localized near a minimum of a
potential well, as shown for the first two gaps by the branches A
and H in Fig. \ref{fig_families} and the corresponding profiles in
Figs. \ref{fig_profiles1_1},  \ref{fig_profiles1_2}, and \ref{fig_profiles2}. Higher-order modes can be thought of
as {\em bound states} of several fundamental modes, including the
continuous counterparts of "twisted modes" of the discrete model
\cite{sukh_pre} and multiple states of the localized modes in nonlinear lattices 
\cite{champ_prb}.

In the first gap (see Figs. \ref{fig_profiles1_1} and \ref{fig_profiles1_2} ), we show only the
lowest-order bound states of two (states B, D, F, and G) or three
(states C and E) in-phase (states D, E, and F) or out-of-phase
(states B, C, and G) fundamental modes. These can be "even", i.e.  centered between potential wells (states
B, C, D, and E) or "odd", i.e. centered on the potential minima (states F and G). For
low condensate densities, the number of atoms in the localized
states that correspond to the bound states of two or three
fundamental modes are, correspondingly, two or three times larger
than that in a fundamental state (see Fig. \ref{fig_families}).
The families of the higher-order formed of the same number of
fundamental states are degenerate in $N$ for low densities; this
degeneracy lifts for larger densities, i.e. for larger mean-field
nonlinearities. It is interesting to note that, along with the
fundamental and its bound states, there exist higher-order localized modes and
their corresponding bound states that {\em do not have analogues}
in the discrete lattice models. A representative mode (K) and its
out-of-phase bound state (L) are shown in Fig.
\ref{fig_profiles2}.

The spatially localized modes exist only in the gaps of the
Bloch-wave spectrum. Closer to the band edges, the mode structure
changes dramatically. In particular, all the localized modes of
BEC with {\em repulsive interaction} ( $\sigma=+1$) develop
extended "staggered" oscillating tails near the upper gap edges,
as demonstrated for the fundamental mode and two of its bound states in
the second gap (Fig. \ref{fig_profiles2}, right column at $\mu
=0.09$). The spatial symmetry and periodicity of these tails
corresponds to the Bloch-wave structure at the corresponding band
edge (see Fig. \ref{fig_profiles2}, second column). Inside the
bands, these localized states (two examples are shown in Fig.
\ref{fig_profiles3}) exist as the modes on an extended,
oscillating background with the structure of the corresponding Bloch wave [cf. Figs. \ref{fig_profiles3} and \ref{pearl_fig0}(c)]. These states, localized on a background, are analogous to the "antidark" optical gap solitons described in Ref. \cite{antidark}, and can, in principle, be  observed experimentally.

\begin{figure}[tbp]
\includegraphics[width=7.5cm]{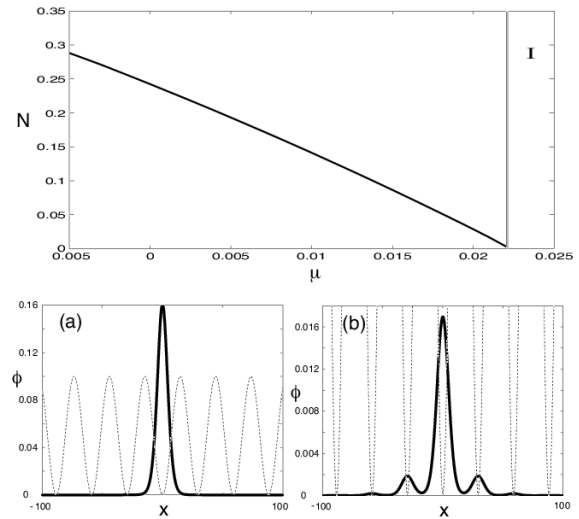}
\caption{\label{attractive_profile}Top: the family diagram and bottom: spatial structure of the lowest-order localized states of the attractive BEC ($\sigma=-1$) in the semi-infinite gap: (a) far from, and (b) near the edge of the first band,  for  $\mu=5 \times 10^{-4}$  and $\mu=2.22 \times 10^{-2}$, respectively.}
\end{figure}

For the {\em attractive interatomic interaction} ($\sigma=-1$),
additionally to the gap modes described above, novel types of localized modes and
their bound states are found in the semi-infinite gap, i.e. below
the lowest edge of the first band of the matter-wave spectrum (see Figs. \ref{fig_stability} and  \ref{pearl_fig0}). The modes of this type, shown in Fig.  \ref{attractive_profile} are
supported by the attractive interaction, and they are similar to
conventional bright solitons with exponentially decaying tails that have no nodes and are weakly modulated by the lattice potential. For any fixed value of the lattice
amplitude, $V_0$, with growing chemical potential, the localized
modes approach the lower edge of the first band of the linear spectrum
and develop oscillating tails with the structure defined by the structure of
the lowest-order Bloch waves [cf. Fig. \ref{pearl_fig0}(a)].

\subsection{Stability of localized modes}

Stability of extended and localized modes in nonlinear systems is
a very important issue, since only dynamically stable modes are likely to be generated
and observed in experiments. To determine the stability properties
of the localized modes,  we consider small perturbations to a
solution of the GP equation (\ref{eq1DGPE}) in the form:
\begin{equation}
\label{perturb}
\psi(x,t)=\phi(x)e^{-i\mu t} + \varepsilon \left[ u(x)e^{i\beta t}
+ w^*(x)e^{-i\beta^*t} \right]e^{-i\mu t},
\end{equation}
where $\varepsilon \ll 1$, and $\phi(x)$ is the steady state
localized nonlinear mode. 
We linearize Eq. (\ref{eq1DGPE}) around the localized solution and obtain, to the first order in
$\varepsilon$, the linear eigenvalue problem for the perturbation
modes:
\begin{equation}
\label{perturb_eig}
\left[ \begin{array}{cc}
\hat{\mathcal{L}}_0 & -\sigma \phi^2  \\
\sigma \phi^{*2} & -\hat{\mathcal{L}}_0
 \end{array} \right]
\left( \begin{array}{c}
u  \\
w \end{array} \right) = \beta \left( \begin{array}{c}
u  \\
w
 \end{array} \right),
 \end{equation}
where
\[\hat{\mathcal{L}}_0 \equiv \frac{1}{2} \frac{d^2}{d x^2} -
V(x)-2\sigma|\phi|^2+\mu.\] In this presentation, the modes describing
the development of instability have either {\em purely imaginary}
or {\em complex} eigenvalues $\beta$; in this latter case the
instability is called {\em an oscillatory instability}. 
We solve the eigenvalue problem (\ref{perturb_eig}) for the
condensate localized states found numerically in the previous
section by standard finite-difference methods or as a matrix eigenvalue problem with Fourier or polynomial interpolants for the differential
operators.

\begin{figure}[htb]
\includegraphics[width=8.5cm]{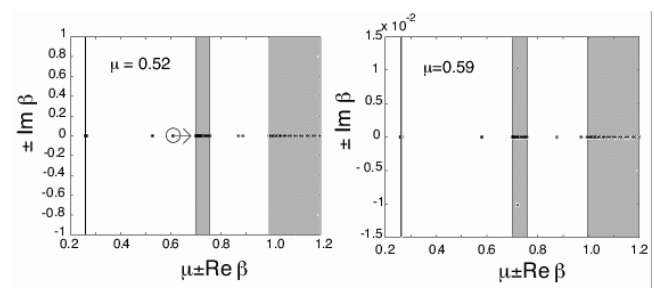}
\caption{\label{pearl_stability_fig0} A typical spectrum of the linear
eigenvalue problem  for the higher-order mode of the repulsive BEC
($V_0=1.0$, $K=0.4$), below (left) and above (right) the
instability threshold. Shaded areas indicate linear Bloch bands.}
\end{figure}

\begin{figure}[htb]
\includegraphics[width=8.5cm]{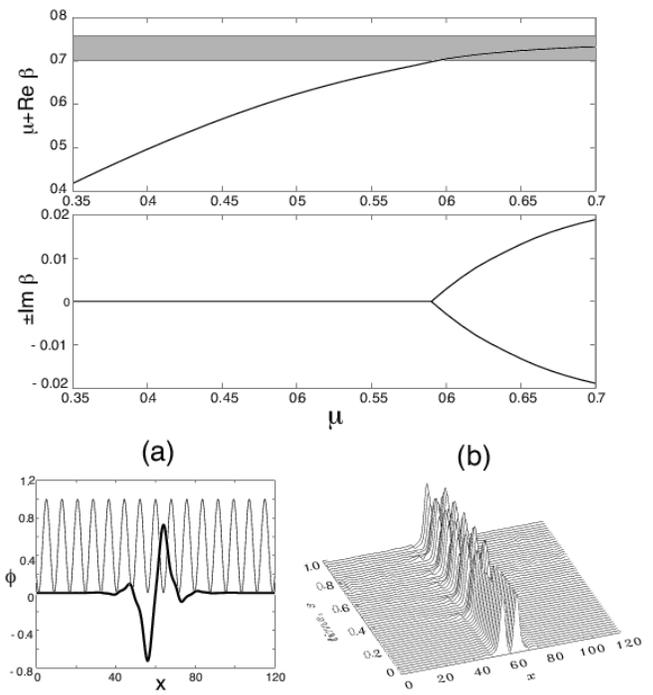}
\caption{\label{pearl_stability_fig1} Top: Example of the
resonance between one of the perturbation modes
and the Bloch-wave band edge leading to an oscillatory instability
of the odd localized mode of the repulsive BEC ($V_0=1.0$,
$K=0.4$).  Bottom: (a) mode profile and (b) its temporal evolution
above the instability threshold 
($\mu=0.65$, $\mu\pm{\rm Re}\beta=0.733$, ${\rm Im}\beta=\pm0.019$).}
\end{figure}

It is well known from the theory of optical gap solitons in nonlinear Bragg gratings, that the linear stability analysis of gap solitons is not trivial and presents significant computational difficulties \cite{Schollmann_1,Barashenkov,Schollmann_2,Schollmann_3}. We have confirmed that the same difficulties arise in the stability analysis of matter-wave gap solitons. The reason for this is the occurence of weak {\em oscillatory instabilities} of gap solitons, which are associated with poorly localized perturbation eigenfunctions. Detection of such instabilities for a {\em finite continuous system}, whether by finite-differences or pseudo-spectral methods, is highly sensitive to the boundary conditions, the size of the spatial domain, and thus to the periodicity of the system. In addition, the discretization of the continuous system of the spatial extent $L$ with $M$ grid points has been shown to produce spurious unstable eigenvalues. In general, the convergence of the perturbation eigenvalues as $\{L, M/L  \} \to \infty$ is exponentially slow \cite{Schollmann_1,Barashenkov}. The effective and accurate detection of oscillatory instabilities is possible for $L\to \infty$, and the infinite domain boundary conditions can be simulated, for example, by expanding the wavefunctions in the basis of Bloch waves (or generalized plane waves in optical case \cite{Schollmann_2,Schollmann_3}). The accuracy of this method is limited by the number of Bloch functions in the expansion. Alternatively, the Evans function method used, e.g., in \cite{sukhorukov,sukh_pre} is capable of treating exact boundary conditions at the edges of the spatial domain.

One should bear in mind, however, that the stability results obtained for the infinite domain may not be confirmed by the dynamical simulations of the time-dependent GP equation due to their essentially finite spatial domains. (For instance, optical gap solitons found to be stable on an infinite domain can be unstable on a finite domain \cite{Barashenkov}).The problem is not only a numerical one since the physical extent of the optical lattices is always limited by the size of the external trapping potential. Below, we present stability analysis for a large but finite lattice without an external potential. The virtue of this analysis is in its qualitative predictions with regard to the types and scenarios of the instability that is exhibited by the localized modes of the BEC in a 1D lattice. Quantitatively accurate, experimentally relevant calculations of both localized states and their stability should ideally treat a finite lattice with the parabolic trapping potential.

A typical structure of the eigenvalue spectrum is presented in
Fig. \ref{pearl_stability_fig0}. Since the perturbation
eigenfunctions are supported by the periodic potential of the
lattice, the full spectrum of the eigenvalue problem  (\ref{perturb_eig}), i.e. the values of $\mu \pm \beta$,  has the band-gap structure identical to that presented in Fig. \ref{fig_stability}. In addition, for any fixed $\mu$, there exist modes localized inside the gaps of the
continuous spectrum, the so-called "internal modes". These linear
gap modes exist due to the distributed "defect" in the lattice
potential introduced by the nonlinear localized states. As the
chemical potential grows, the internal modes of the localized
states move towards the upper edges of the continuum bands (see Fig.
\ref{pearl_stability_fig0}) and generate an {\em oscillatory
instability} through the resonance mechanism previously described
 in a different physical context \cite{sukhorukov}. Because different "internal"
perturbation modes with different symmetries can cross into the
band and resonate with the linear Bloch waves, the instability can
trigger various types of spatial dynamics, including the symmetry
breaking instability.

Applying the stability analysis to the modes identified in Sec. IV A, we
found that {\em all of the higher-order gap modes}
identified for our model exhibit {\em oscillatory instability} in
the regions of the existence domain near the upper band edges. For the parameters used in our calculations the strong exponential instability associated with a purely imaginary $\beta$ {\em does not occur} has not been detected for gap modes. A typical evolution of the internal perturbation mode in a gap and birth of the oscillatory instability for large $\mu$ is shown in Fig. \ref{pearl_stability_fig1} (top), for the even, out-of-phase gap mode. The time scale for development of such weak  instabilities is large, which means that oscillatory unstable modes can be dynamically stable  on the scales comparable with the typical lifetime of the condensate. 

For the lowest-order fundamental gap modes, our analysis has shown the existence of internal modes which, if excited, may be responsible for long-lived oscillations. It also produced very narrow regions of extremely weak oscillatory instability (${\rm Im} \beta\sim~10^{-4} - 10^{-5}$)
in the close vicinity of the band edges. The accuracy of these calculations is low since the gap mode is poorly localized in this region. As the computed values ${\rm Im} \beta \to 0$ with increased accuracy of the method, it is most likely that this oscillatory instability is {\em unphysical}   and the fundamental gap solitons are {\em linearly stable}. 

\begin{figure}[htb]
\includegraphics[width=8.5cm]{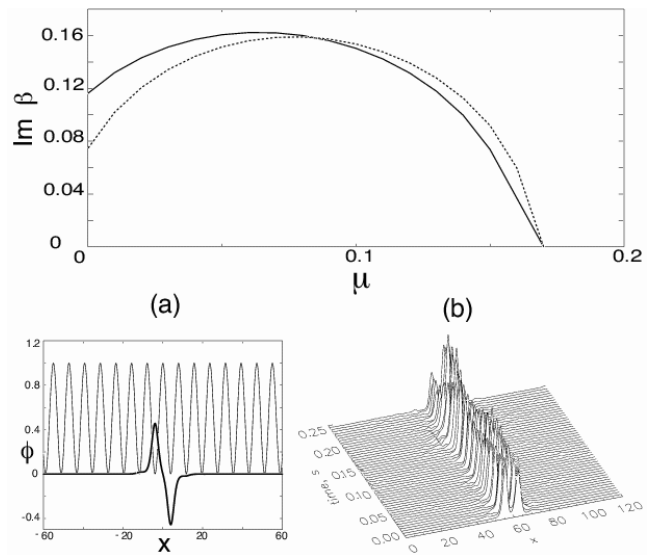}
\caption{\label{attr_growthrate} Top: Growth rate of
the strong linear instability for an even out-of-phase localized mode of the {\em attractive} BEC ($V_0=1.0$,
$K=0.4$, $\sigma=-1$); dotted and solid lines correspond to the double-peaked even in- and out-of-phase perturbation modes, respectively. Bottom: (a) the mode profile and (b) its temporal evolution; the mode is initially perturbed by off-setting the position of its center  ($\mu=0.1$, $\beta=0.157i$).}
\end{figure}

In contrast, some of the higher-order localized modes of the {\em attractive
condensate} ($\sigma=-1$), in the semi-infinite gap below the first band,  have been found to experience the strong (exponentially growing) linear
instability, characterized by a pair of purely imaginary
eigenvalues of the perturbation modes $\beta$. These modes originate from neutral modes of the excitation spectrum (with ${\rm Re} \beta =0$). An example of such
instability development via bifurcation of the neutral mode into the imaginary plane is presented in Fig. \ref{attr_growthrate} for an odd bound state of fundamental bright solitons. Remarkably, the instability growth rate drops as the mode moves away from the lowest band edge. This is due to the fact that the condensate density grows and, ultimately, strong localization of large-amplitude BEC droplets occurs at the individual wells. In this limit the lattice effects are weak, and the property of a localized state is described by the free-space GP equation, for which the (inverted for our choice of $\mu$) {\em Vakhitov-Kolokolov stability criterion} states that the fundamental bright solitons of attractive BEC are stable as long as $-\partial N/\partial \mu>0$, i.e. in the entire semi-infinite gap (see Fig. \ref{attractive_profile}, top).

In order to check the results of the stability analysis of the
stationary solutions for the BEC modes, we have employed direct
numerical simulations of the time-dependent equation
(\ref{eq1DGPE}), using a split-step Fourier method which preserves
the normalization of the condensate wave function with a high
accuracy. To confirm the stable dynamics of gap solitons, the results were obtained by running the
simulations for 8000 dimensionless time units which is equivalent
to a time of 10.6 seconds.  Barrett {\em et al.} \cite{Bar_et01}
who created the first BEC directly in an optical trap, found that
for a condensate of $3.5\times 10^{4}$ atoms the $1/e$ lifetime
was about 3.5 seconds. However, Stamper-Kurn \textit{et
al.}~\cite{Sta_et98} observed that in a purely optical trap the
$1/e$ lifetime at low densities was longer than $10$ ${\rm s}$, although
at that timescale significant loss occurs as the population goes
from $10^{7}$ atoms to $10^{5}$ atoms. The results of our
dynamical simulations showing the different instability
developments for oscillatory and linearly unstable higher-order
states of repulsive and attractive condensates are presented in Figs.
\ref{pearl_stability_fig1}(b) and \ref{attr_growthrate}(b),
respectively.

\section{Conclusions}

We have analyzed the stationary states and stability of the
condensate in an optical lattice, in the framework of the GP
equation with a periodic potential. We have confirmed some of the
earlier predictions for the condensate structure based on the
effective tight-binding model, as well as revealed a number of
novel features not described by the effective discrete models. In
particular, we have extended the concept of the Bloch waves to the
case of a nonlinear model and described the families of the
nonlinear periodic states which generalize the corresponding
Bloch-wave modes of linear periodic systems. We have demonstrated
that the spectrum of such periodic modes possesses a band-gap
structure, which depends on the number of atoms, i.e. the
effective nonlinearity in the mean-field theory.

In the gaps of the matter-wave spectrum, we have numerically found
the families of novel spatially localized modes of the condensate
in an optical lattice trapped by a few lattice minima. Such modes
should be compared with optical gap solitons, and they can exist even for
the condensates with repulsive interaction. We have analyzed the
structure and, more importantly, the linear stability of the gap
solitons which belong to different band gaps, for both repulsive
and attractive interatomic interactions. We believe that these
novel types of spatially localized, coherent excitations of the
condensate can potentially be observed in experiments.

Our studies extend the analysis of the BEC in an optical lattice beyond 
the applicability limits of the frequently used
discrete models based on the tight-binding approximation, as well as they call for the future
extensive analysis of the condensate stability in optical lattices
described by the three-dimensional GP equation with a periodic potential. In
particular, similar to the condensates in a parabolic trap, we
expect that optical lattices can support dark solitons, excited on
the different types of the nonlinear Bloch waves. The stability of
such dark solitons should differ dramatically from the stability
of the conventional dark solitons. These studies are now in
progress.

\section*{Acknowledgements}

We are indebted to N. Robins for help with numerical
simulations and to A.A. Sukhorukov for critical reading of this manuscript. We thank C.J. Williams and  P.V. Elyutin for useful
discussions and suggestions. This work was partially supported by the
Australian Research Council and the Australian Partnership for
Advanced Computing.

\bibliography{1d_lattice}

\end{document}